\documentstyle[11pt,hrd_pasp,twoside,epsf]{article}
\begin{document}
\title{The Ca II Triplet as an Abundance Indicator}
\author{Bosler, T.}
\affil{Department of Physics \& Astronomy,
4129 Fredrick Reines Hall,
University of California,
Irvine, CA 92697$-$4675}
\author{Smecker$-$Hane, T.}
\affil{Department of Physics \& Astronomy,
4129 Fredrick Reines Hall,
University of California,
Irvine, CA 92697--4675}
\author{McWilliam, A.}
\affil{Observatories of the Carnegie Institute of Washington;
  813 Santa Barbara St., Pasadena, CA 91101}

\begin{abstract}
How can we accurately determine the metallicity of faint
red giant stars in nearby galaxies?  The equivalent widths of the broad
absorption lines
from Ca II at 8498 \AA, 8542 \AA, and 8662 \AA~ (the calcium triplet) 
have shown a smooth relationship to metallicity, [Fe/H], in Galactic
globular clusters, i.e., old, metal$-$poor stars.
The exact relationship depends on the [Fe/H] scale assumed.
The sensitivity of the sum of the equivalent widths, $\sum{W_{Ca II}}$, to 
[Fe/H] has been well studied for metal$-$poor stars, but theory and
observations show that $\sum{W_{Ca II}}$, will become less sensitive to [Fe/H] 
for young or metal$-$rich stars, but this needs to be calibrated.
Built into the present calibration of $\sum{W_{Ca II}}$ to [Fe/H] is
also a dependence on the Galactic [Ca/Fe] to [Fe/H] relationship,
which is a function of the Galaxy's star formation history (SFH)
and chemical evolution.
Our goal is to remove uncertainties in the [Fe/H] scale used in previous
calibrations and remove the dependence on galactic evolution.  We are obtaining 
high dispersion spectra to self-consistently calibrate the calcium triplet to 
give [Ca/H] abundances, and extend the current calibration to near$-$solar 
metallicities and to ages as young as $\sim$ 2 Gyrs.
\end{abstract}
\section{Introduction}
Establishing the absolute and relative ages of stars is a
problem which continues to plague modern astronomy.  Obviously,
the accurate determination of these ages
places strong constraints on the chemical evolution and SFH of the galaxy in
which the stars reside.  Photometry alone cannot be used to determine age 
because of the age$-$metallicity degeneracy of broad--band colors.
High dispersion observations can accurately determine metallicity and, combined 
with the CMDs, tie down ages.
 However, high--dispersion observations require long exposure times to obtain a
high signal--to--noise ratio (S/N) and, for faint or extra--galactic stars, this 
becomes impractical.  To
determine cluster ages, methods to find metallicity accurately from other 
observational techniques are needed.

The near$-$infrared Ca II triplet, at 8498 \AA, 8542 \AA, and 8662
\AA, are very strong spectral features (See Fig. 1).
With moderate S/N ($\approx$ 50) and
moderate instrumental resolution ($\approx$ 4 \AA), the sum of the equivalent 
widths, $\sum{W_{Ca II}}$, can be determined with an accuracy of $\sigma_{W_{Ca 
II}}$ $\approx$ 0.10 (see, for example, Rutledge, et al.~ [1997a]; hereafter, 
Rut97a) making it a useful tool if its relationship to metallicity is well 
determined.
Zinn \& West [1984] (ZW84) found a smooth correlation between
$\sum{W_{Ca II}}$ and iron abundance, [Fe/H].  Their [Fe/H] scale was
developed from a large data set of 121 Galactic globular clusters, GGCs, with 
[Fe/H] determined from various spectroscopic indices.
Armandroff \& Da Costa [1991] found that $\sum{W_{Ca II}}$ changes with V $-$
V$_{HB}$ due
to changes in gravity.  To remove the effects of gravity to first
order, they defined a new Ca II index, $W^\prime =
\sum{W_{Ca II}} - 0.62 $(V $-$ V$_{HB})$ where $\sum{W_{Ca II}}$ is the sum of
the equivalent widths of
the two strongest Ca II lines.  Carretta and Gratton [1997] (CG97) rederived the
calibration with
a much smaller data set than ZW84, 24 GGCs, but determined [Fe/H] values using
only
high dispersion spectroscopy of Fe lines.  The CG97 and ZW84 scales yield
different relationships for $\sum{W_{Ca II}}$ to [Fe/H].  To clarify the
differences, Rutledge, 
et al.~[1997b] (Rut97b) systematically compared the [Fe/H] scales used by the 
two groups.

Rut97a observed 52 GGCs to determine the Ca II index and used these
values to compare the ZW84
and CG97 [Fe/H] scales.  The Ca II index developed by Rut97b is $W^\prime =
\sum{Ca II} + 0.64$ (V $-$ V$_{HB}$)
where $\sum{Ca II} = 0.5~W_{8498} + W_{8542} + 0.6~W_{8662}$ (with weights
chosen to minimize
$\sigma_{Ca II}$).  They found that the $W^\prime$
to [Fe/H] relation was non-linear for the ZW84 scale and linear for the CG97
scale.  In
fact, the differences in the derived values of [Fe/H] can be as great as 0.3 dex
depending
upon the [Fe/H] scale used.  This can lead to errors of several Gyrs when using
these
metallicities to determine clusters ages.  The differences in the scales are
most dramatic
for young and metal$-$rich clusters which were not used to determine either
calibration.  Some
theoretical work (e.g., Jorgensen, et al.~[1992]) has shown that the
dependence of $W^\prime$ on [Fe/H] may change for
more metal$-$rich stars so the relationship will change at high [Fe/H].  The
uncertainty in
the $W^\prime$ to [Fe/H] relationship over a range of age and metallicities
remains undetermined though the ZW84 and the
CG97 relationships have been widely used.

An additional uncertainty in the calibration of $W^\prime$
to [Fe/H] is the assumption of a Galactic [Ca/Fe] vs.~[Fe/H] relationship.  
All observations used to determine the scales were
done using GGCs which have a [Ca/Fe] vs. [Fe/H] relationship that depend on the 
unique chemical evolution and SFH of our Galaxy.  Also, considerable
scatter around the mean relationship exists.  Since Ca comes primarily
from Type II supernovae and Fe comes from both Type II and Type Ia
supernovae (e.g., Carney, B.W.~[1996]), the ratio
of these elements will depend upon the SFH and chemical evolution of the Galaxy.
Applying the
relationship to extra$-$galactic stars adds an uncertainty to the derived 
abundances.

To properly calibrate the relation of $W^\prime$ to abundances and
remove the dependence on SFH, we propose to calibrate the relationship between
$W^\prime$ and [Ca/H].  In addition, we will recalibrate and extend the
calibration of $W^\prime$
to our new [Fe/H] scale.  $W^\prime$ will be determined from low dispersion
spectra and our results tied to the well-defined scaled of Rut97a.  [Ca/H] and 
[Fe/H] will be determined by
measuring Fe I, Fe II and Ca I from newly obtained, high--dispersion spectra.
We will be observing Galactic open and globular cluster stars in order to assure 
that our new calibration is valid over a wide range of ages and abundances.
\section{Observations}
We will be using observations of 25 star clusters to
determine the relation of $W^\prime$ to [Ca/H] and $W^\prime$ to
[Fe/H]. Twenty are globular clusters with approximate ages of 14
Gyrs and iron abundances $-2.40 \le$ [Fe/H] $\le -0.73$; most were well studied
by Rut97
and have well determined Ca II strengths ($\sigma_{W^\prime} \le 0.10$) but
[Ca/H] and [Fe/H] need to be determined.  Five clusters are open clusters with
ages from 1.6 to 8 Gyrs and -0.4 $\le$ [Fe/H] $\le$ 0.4.
 Open clusters have not been as well studied spectroscopically so we will
measure $W^\prime$, [Ca/H] and [Fe/H].
\subsection{Low Dispersion}
$W^\prime$ measurements were made on spectra taken with
the Kast Double Spectrograph on the Shane 3--meter Telescope at the
University of California's Lick Observatory. A dispersion of 1.7 \AA~ pix$^{-1}$
and a wavelength coverage of 7580 $-$ 9620 \AA~
(resolution $\approx$ 4 \AA~) were used for the low dispersion
observations.  Exposure times were chosen to give S/N $\approx$
50, which will yield $\sigma_{W^\prime} \approx$ 0.1 \AA~ per.

To assure accuracy in our values of $W^\prime$, we
average over 10 to 15 stars per cluster which were selected from
 broad--band photometry and radial velocity data, if available.
\subsection{High Dispersion}
Most of our high dispersion observations were made using the
Hamilton$-$Echelle
Spectrograph on the Shane 3--meter Telescope at Lick Observatory.
Four of the fainter clusters were observed with
the High Resolution Spectrograph on the Keck I 10--meter Telescope at the W.M.
Keck Observatory.  These observations were made
with a dispersion of 0.05 \AA~ pix$^{-1}$  and a wavelength coverage of
5200 $-$ 9000 \AA~ (resolution $\approx$ 0.15 \AA~).   Exposure times were 
chosen to give S/N $\approx$ 60 per star.  Averaging over 5 to 8 red giants per 
cluster we should obtain abundances with $\sigma_{[Ca/H]}$ and $\sigma_{[Fe/H]} 
\approx$ 0.1 dex per cluster.

The large wavelength range is necessary since we will be measuring many
lines from these spectra in order to determine [Ca/H] and [Fe/H]: 10 to 15 Ca I 
lines; $\sim$ 70 Fe I lines; 4 to 8 Fe II lines.  Measurements of the equivalent 
widths will be made using the GETJOB program  (A. McWilliam, 1999).  Ca I, 
rather than Ca II, lines will be used to determine [Ca/H] because the formation 
of the  weak Ca I lines are well understood and
accurately determined using local thermodynamic equilibrium (LTE) models while
the formation of the strong Ca II lines are more complex.
 
{\bf Acknowledgements:} Financial support for this project
was provided by National Science Foundation grant AST-0070985
to TSH. TB also thanks the ARCS Foundation for fellowship
support.

\begin{figure}
\vspace{-1.0cm}
\plotfiddle{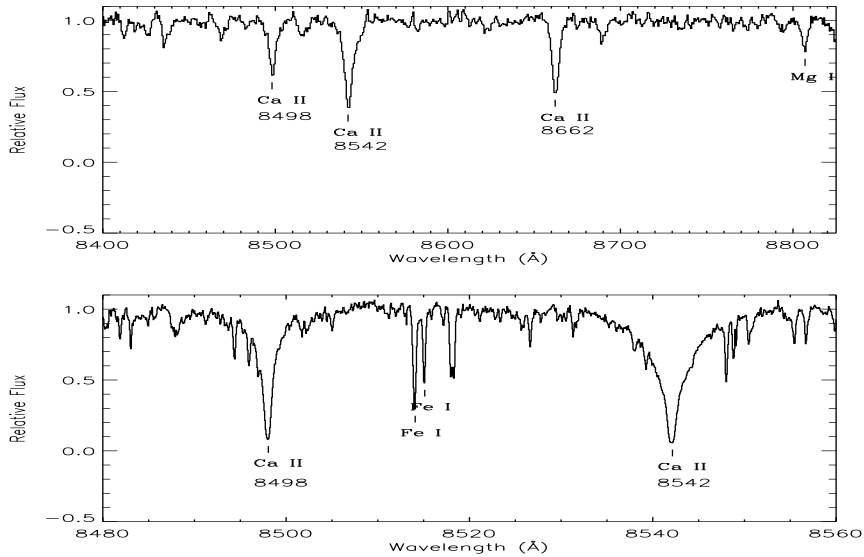}{8.0cm}{0.}{60.}{30.}{-180}{-10}
\caption{Low dispersion (Top) and high dispersion (Bottom) spectra$-$
note different $\lambda$ range.  Notice the how the broad wings of the strong
Ca II lines are apparent in the high dispersion spectra.  The Fe I (and Ca I)
lines are so weak that they cannot be resolved accurately at low dispersion.}
\end{figure}

\end{document}